\documentclass[namedreferences]{solarphysics}
\usepackage[hyperref,optionalrh]{spr-sola-addons} % For Solar Physics 
\usepackage{graphicx}        % For eps figures, newer & more powerful
\usepackage{url}        % For breaking URLs easily trough lines
\usepackage[usenames,dvipsnames]{color} % For color text: \color command
\usepackage[normalem]{ulem}     %Provides the following commands:
\usepackage{pdflscape}

\newcommand{\BE}{\begin{equation}}
\newcommand{\EE}{\end{equation}}
\newcommand{\BA}{\begin{eqnarray}}
\newcommand{\EA}{\end{eqnarray}}

\newcommand{\degree}{^\circ} 
\newcommand{\ms}{m s$^{-1}$ }

\chardef\us=`\_

\begin{document}

\begin{article}
\begin{opening}

\title{Variation of Chromospheric Features as a Function of Latitude and Time using Ca-K Spectroheliograms for Solar Cycles 15 -- 23: Implications for Meridional Flow}
\author[addressref={aff1},corref,email={setiapooja.ps@gmail.com}]{\inits{Pooja}\fnm{Pooja}~\lnm{Devi}\orcid{https://orcid.org/0000-0003-0713-0329}}%\sep
\author[addressref={aff2}]{\inits{Jagdev}\fnm{Jagdev}~\lnm{Singh}}%\sep
\author[addressref=aff1]{\inits{Ramesh}\fnm{Ramesh}~\lnm{Chandra}\orcid{https://orcid.org/0000-0002-3518-5856}}%\sep
\author[addressref=aff2]{\inits{Muthu}\fnm{Muthu}~\lnm{Priyal}}%\sep
\author[addressref=aff1]{\inits{Reetika}\fnm{Reetika}~\lnm{Joshi}\orcid{https://orcid.org/0000-0003-0020-5754}}%\sep

\address[id=aff1]{Department of Physics, DSB Campus, Kumaun University, Nainital 263 002, India}
\address[id=aff2]{Indian Institute of Astrophysics, Bangalore- 560 0034, India}

\runningauthor{P. Devi {\it et al.}}
\runningtitle{Meridional Flow During Solar Cycles 15 -- 23}

\begin{abstract}
We have analysed the Ca-K images obtained at Kodaikanal Observatory as a function of latitude and time for the period of 1913 -- 2004 covering the Solar Cycle 15 to 23. We have classified the chromospheric activity into plage, Enhanced Network (EN), Active Network (AN), and Quiet Network (QN) areas to differentiate between large strong active and small weak active regions. The strong active regions represent toroidal and weak active regions poloidal component of the magnetic field. We find that plages areas mostly up to 50$\degree$ latitude belt vary with about 11-year Solar Cycle. We also find that weak activity represented by EN, AN and QN varies with about 11-year with significant amplitude up to about 50$\degree$ latitude in both the hemispheres. The amplitude of variation is minimum around 50$\degree$ latitude and again increases by small amount in the polar region. In addition, the plots of plages, EN, AN and QN as a function of time indicate the maximum of activity at different latitude occur at different epoch. To determine the phase difference for the different latitude belts, we have computed the cross-correlation coefficients of other latitude belts with 35$\degree$ latitude belt. We find that activity shifts from mid-latitude belts towards equatorial belts at fast speed at the beginning of Solar Cycle and at slower speed as the cycle progresses. The speed of shift varies between $\approx$ 19 and 3 m s$^{-1}$ considering all the data for the observed period. This speed can be linked with speed of meridional flows those believed to occur between convection zone and the surface of the Sun.
\end{abstract}

\end{opening}
%-------------------------------------------------

%%%%%%%%%%%%%%%%%%%%%%%%%%%%%%%%%%%%%%%%%%%%%%%%%%%%%%%%%%%%%%%%%%%%%%%%%%%%%%%%%%%%%%
\section{Introduction}
\label{sect_Introduction}
%%%%%%%%%%%%%%%%%%%%%%%%%%%%%%%%%%%%%%%%%%%%%%%%%%%%%%%%%%%%%%%%%%%%%%%%%%%%%%%%%%%%%%

Solar surface depicts various activity features that vary on short and long time scales. The continuum images of the Sun show sunspots and white light faculae. Sun's images in X-rays and EUV wavelength domain show bright points loop structures in emission. The historic images of the Sun obtained in H$\alpha$ and Ca-K lines show plages, filaments and networks. The magnetograms of the Sun show high magnetic field at the location of sunspots and plage regions. These also indicate weak magnetic field spread over the whole visible surface of the Sun. All these features are related to each other and their occurrence varies with time. The Ca-K images show plages, enhanced network (EN), Active network (AN) and quiet network (QN). These represent magnetic field regions of different strength and their brightness varies accordingly. \inlinecite{Leighton1959} found that the Ca-K plage regions have magnetic fields in the range of 100 -- 200 Gauss. \inlinecite{Simon1964} showed a strong correlation between the boundaries of the large convective cells known as super-granules and Ca-K network. These magnetic features have cyclic variations (\opencite{Oliver1998, Gopalswamy2008} and references cited therein). From the spectroscopic observations of the Ca-K profiles, \inlinecite{Sindhuja2014} reported that variations in the Ca-K line parameters can be used to study the meridional flows. The movement of solar activity towards lower latitudes has been studied by using sunspots and related high magnetic field indices. During the later part of the 20$^{th}$ century observations were made to make magnetic field images of the Sun. Using these images pole and equator ward shift in the activity, thereby the meridional flows have been studied. The magnetic images of the Sun for the earlier times are not available but other type such as Ca-K, H$\alpha$ are available since the beginning of 20$^{th}$ century (\opencite{Oliver1998, Gopalswamy2008, Sindhuja2014,Priyal2019} and references cited therein). 
The cyclic variations of these features are known as Solar Cycle. The Solar Cycle is known to us since one and half century and discovered by \inlinecite{Schwabe1843} at first. After that, \inlinecite{Hale1908} showed that sunspots are strongly magnetized. Hale's observations revealed that the complete magnetic cycle spans two Solar Cycles {\it i.e.} 22 years, before returning to its original state, called solar magnetic cycle. This cyclic variation of the solar magnetic cycle is due to the cycle continuously changing from toroidal to  poloidal magnetic field then again from poloidal to toroidal magnetic field.  They can interchange from one another as follows: The poloidal field is stretched by the solar differential rotation and as result of this the toroidal field is generated. Again it converts to a poloidal field  by the helical turbulence. 

The flow of plasma material towards or from the equator is known as meridional flow. It is believed that the polar magnetic field is generated by the movement of weak magnetic field towards the 
poles from the decaying active regions \citep{Howard1981}. \inlinecite{Babcock1961} first speculated that the meridional flows is a main transport process to explain the observed pole-ward motion of flux from the sunspot belts. In addition to this the meridional flow towards the equator plays an important 
role for the transportation of the toroidal component of the magnetic field.
The meridional flow is one of the important and crucial results of the dynamo models and can be studied from the shift in the occurrence of the active regions on the Sun \citep{Choudhuri1995, Schrijver2002, Hathaway2003, Baumann2006, Jiang2009, Jin2012}.
\inlinecite{Jiang2009} found in their model, the decrease in polar field and they explained it as a result of
equatorial meridional flow. They reported the speed is of the order of few \ms.

There are observational evidences of meridional flow since long time and it is of great importance for the advancement
of solar dynamo models. The pole-ward transportation of magnetic flux was first confirmed by \inlinecite{Howard1974}.
Lateron it is confirmed in more observational studies \citep{Svalgaard1978,Wang1988,Cameron1998,Durrant2004}.
Using the vector spectromagnetograph data of Synoptic Optical Long-term
Investigations of the Sun (SOLIS) in chromospheric line \inlinecite{Raoufi2007} explored the
distribution of magnetic flux elements as a function of latitude in polar solar caps during minimum of Solar Cycle 23. Their results shows the flux transport towards pole and also reported that 
the meridional circulation responsible for the flux transport slows down before reaching to pole.
The equatorial and pole-ward transport of the flux is recently presented in the study of \inlinecite{Sindhuja2014}. 
They have used the Kodaikanal Ca-K line spectra data for the Solar Cycle 22 -- 23. They found the Ca-K$_1$ width attains
maximum amplitudes at different latitude belts at different phases of Solar Cycle.  
 
Since there is a strong correlation between the magnetic field and the 
Ca-K emission \citep{Skumanich1975,Nindos1998}, the Ca-K line observations can be taken as the proxy 
for the magnetic field to study the meridional flow. 
With this background, we have studied the variations in plage, EN, AN and 
QN areas representing large scale and small scale magnetic activity on the Sun, as a function of time and latitude. The plages represent large scale magnetic activity and mostly can be seen from equator to mid-latitudes. Whereas networks represent small scale activity and can be seen over the whole of visible surface.
The period of the study is for the Solar Cycle 15 -- 23.
The paper is organised as follows: In section~\ref{sect_Observations}, we present the observational data 
sets and data reduction technique. The results about their implications on the meridional flows are discussed in Section~\ref{sect_results}. Finally, the summary
is presented in Section~\ref{sect_Summary}.

%%%%%%%%%%%%%%%%%%%%%%%%%%%%%%%%%%%%%%%%%%%%%%%%%%%%%%%%%%%%%%%%%%%%%%%%%%%%%%%%%%%%%%
\section{Data and Analysis}
\label{sect_Observations}
The details of the data and its digitization are described in a paper by \inlinecite{Priyal2014} and results of investigation of variations in the intensity of above mentioned features are reported by \inlinecite{Priyal2017}. The high spatial resolution (0.86 arcsec pixel$^{-1}$) and 16-bit read out with high photometric accuracy of digitized data permits to identify the network elements and study their variation with time. They found some scatter in the results and careful analysis of the data indicated short and long term variations in the quality of Ca-K obtained over a century. \inlinecite{Singh2018} shorted out the data in two-time series termed as ``Good'' consisting of uniform images and other as ``Okay'' having the remaining images. 
The ``Good'' series has relatively uniform images in terms of
contrast and without any defects due to developments or passing clouds during observations, obtained at Kodaikanal observatory on daily basis for the period of 1907 -- 2004. 
About two-third of the total images come under the category of ``Good'' time-series. \inlinecite{Priyal2019} determined the threshold value of intensity and area to determine different features mentioned above. But the threshold values for the study of variations as
a function of latitude are different than those for the whole image, since some part of
plages and networks are likely to fall in two latitude belts. We found that intensity $>$ 1.30
and area threshold of 0.1 arcmin$^2$ for the plages, intensity $>$ 1.30 and area threshold of 4
arcsec$^2$ for the EN, intensity $<$ 1.30 and $>$ 1.20 and area threshold of
0.1 arcsec$^2$ for the AN and intensity $<$ 1.20 and $>$ 1.10 and area threshold of 4
arcsec$^2$ for QN are suitable to identify these features. The limiting value
of 4 arcsec$^2$ has been chosen to avoid the noise in the data due to 1 or 2 pixels having
values $>$ 1.10 due to photon noise. The image is converted in the binary format
depending on intensity threshold values for the plages, EN, AN and QN. Then the image
is subjected to identify the groups of consecutive pixels representing plages or EN or
AN or QN. If the area limits are satisfied, then those pixels are saved with the actual
intensity values and the other pixels are made as zero. Thus, the final image has the
extracted active regions within the threshold intensity values and area limits. After the
identification of these features, Ca-K line image is divided in heliographic latitude zones
with 10$^\circ$ interval up to 70$^\circ$ north and south latitudes considering the date of
observations and size of the image and 70$^\circ$ to solar limb is considered as one latitude
zone, which has seasonal variations. Then we determine the fractional area and average
intensities of plage, EN, AN and QN for each latitude zone as defined above on daily
basis.

With an aim to determine the long-term variations in these features and to reduce the scatter in the data, we have taken monthly averages. Then the running average of this monthly averaged data over one year upto 40$^\circ$ latitude and three years for latitude belts $>$ 40$^\circ$ is taken. Further, we have computed the cross-correlation function between data of different latitudes to determine the phase differences between activity at various latitudes. 

%%%%%%%%%%%%%%%%%%%%%%%%%%%%%%%%%%%%%%%%%%%%%%%%%%%%%%%%%%%%%%%%%%%%%%%%%%%%%%%%%%%%%%

%%%%%%%%%%%%%%%%%%%%%%%%%%%%%%%%%%%%%%%%%%%%%%%%%%%%%%%%%%%%%%%%%%%%%%%%%%%%%%%%%%%%%%
\section{Results and Discussion}
\label{sect_results}

\begin{figure*}
\centering
\vspace{-0.1cm}
\includegraphics[width=10cm]{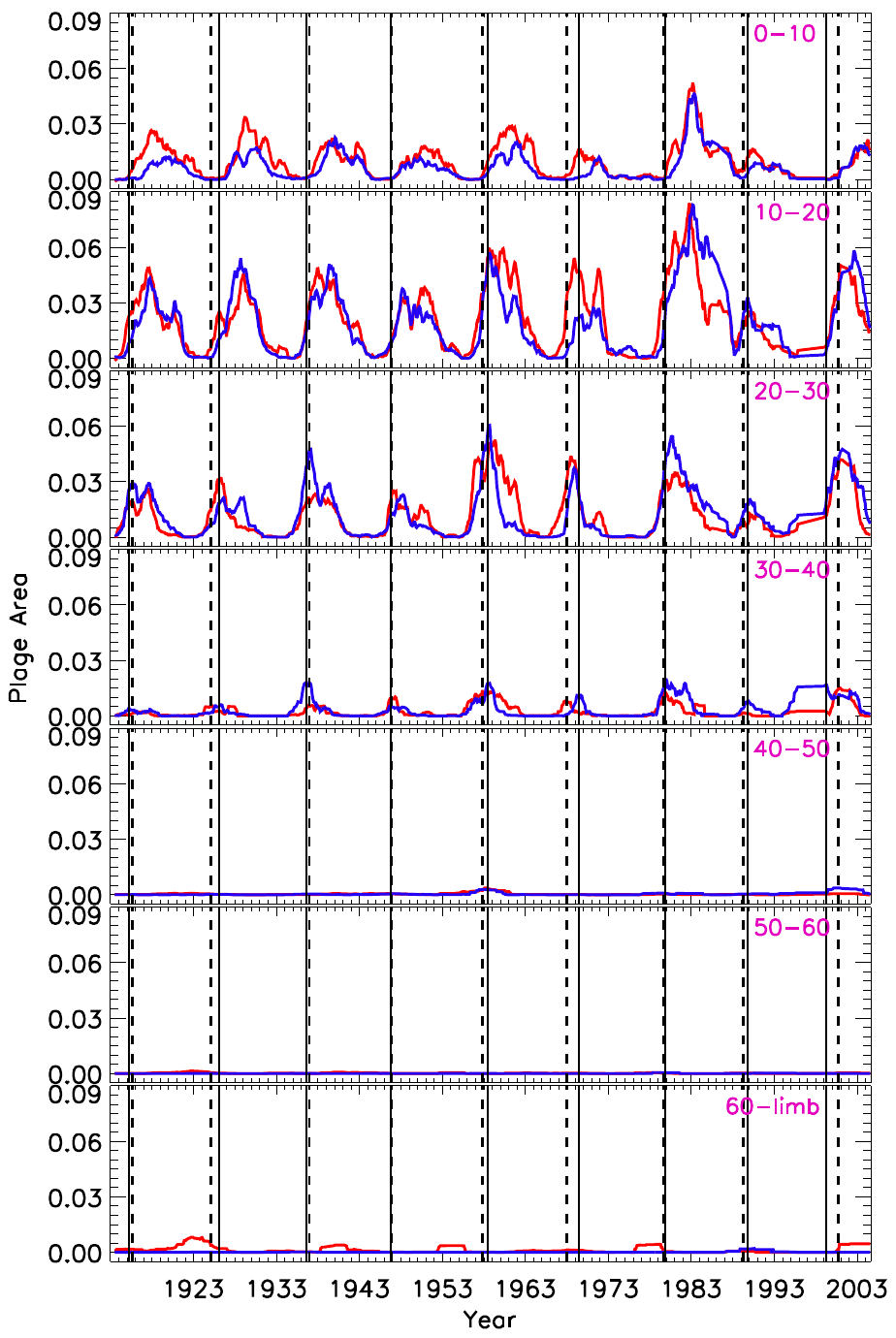}
\caption{Plots of averaged fractional plage area as a function of time for the period of 1913 -- 2004 for different latitude belts in northern (red) and southern (blue) hemisphere. The latitude belt is written on top right of each panel. The vertical dashed and solid lines corresponds to approximate beginning of large activity at 30 -- 40$\degree$ latitude belt for each Solar Cycle in northern and southern hemispheres, respectively.}
\label{p_north}
\end{figure*}

\subsection{Variation of Ca-K Features}
The study of long-term variations in solar surface features is important to understand the solar dynamo process. 
We have segregated the Ca-K active features in four types as classified by \inlinecite{Worden1998}. The Ca-K plages are the relatively bright and large regions generally occur around the sunspot regions. It is believed that decaying Ca-K plages fragments into small regions termed as EN and then decay. The AN is probably related with the small-scale active regions, also known as Ca-K bright points. The QN generally represents the boundaries of the Ca-K network (super-granular network). \inlinecite{Leighton1962} and \inlinecite{Simon1964} have shown a strong correlation between large convective cells and Ca-K network. Thus, plages represent toroidal and networks represent poloidal fields on the Sun. The plages, EN, AN and QN represent different physical characteristics of the chromosphere and thus may show different behaviour. Some features may show variations related with the phase of the Solar Cycle and some may not show. There may be a phase-lag in the variations of different features. With this view, we have studied their variation with time. Thus, the variations in the chromospheric features such as plage, EN, AN, and QN areas with time provide a valuable tool to study the shift in activity from one latitude to other latitude and thereby the meridional flows. In principle, butterfly diagram, generally made using the occurrence of plages with time provide such information but up to middle latitudes only. To determine velocity of shift one needs the detailed analysis of the data. The detection of these features has been shown in Figure 8 of the paper by \inlinecite{Priyal2014}. Here, we determine the fractional areas of these chromospheric features at different latitude belts at an interval of 10$\degree$ for the northern and southern hemispheres. Fractional areas are defined as the areas occupied by the selected feature in pixels divided by the total number of pixels in considered latitude belt.

\subsubsection{Variation in Plage Areas with Latitude and Time}
Ca-K plages regions are large areas around the sunspots with magnetic field in the range of 100 -- 200 Gauss. The variations in the plage areas indicate the variation in the toroidial magnetic field on the Sun. In Figure \ref{p_north} (upper six panels), we plot the monthly averaged fractional plage areas for latitude belts up to 60$\degree$ at an interval of 10$\degree$ for the northern and southern hemispheres. The red and blue curves indicate the variation of plage areas in the northern and southern hemispheres, respectively. The area of the latitude belts greater than 60$\degree$ is small due to projection effect. Therefore, in the bottom most panel of the figure we show the monthly averaged plage area as a function of time for the latitude belt of 60$\degree$ -- limb (visible part). The polar region covered varies depending on B$_\circ$ angle of the image of the Sun. The figure indicate activity due to plages at 30 -- 40$\degree$ belt remain for short duration as compared to lower latitude belts between equator to 30$\degree$ in both the hemispheres. A look at the figures indicate that fractional plage areas mostly occur up to 40$\degree$ latitude belt in both the hemispheres. The fractional plage areas at higher latitude belts appear to be negligible as expected. The figure also indicate that the plages occur more in latitude belt of 10 -- 20$\degree$ both in the northern and southern hemispheres as compared to other latitude belts. This is consistent with the findings of \inlinecite{Priyal2017}. The Figure 6 of their paper indicate that the number of plages peaks around 10 -- 20$\degree$ latitudes for each Solar Cycle from 14 -- 23. The figure also shows that on an average plages vary with 11-year period. The Solar Cycle begins around middle latitude belts and active regions occur at lower and lower latitudes as the cycle progresses. We, therefore, have found the approximate epoch of the large scale occurrence of plages at the latitude belt of 30 -- 40$\degree$ for each Solar Cycle and shown by dotted black line in the figures. The epoch of maximum activity represented by fraction plage area at other latitude belts does not coincide with the peak of 30 -- 40$\degree$ latitude belt in both the hemispheres. Using the cross-correlation function we have computed the phase differences between the occurrence of maximum activity at various latitude belts with respect to 30 -- 40$\degree$ latitude belt which will be discussed in this paper later.

\begin{figure*}
\centering
\includegraphics[width=10cm]{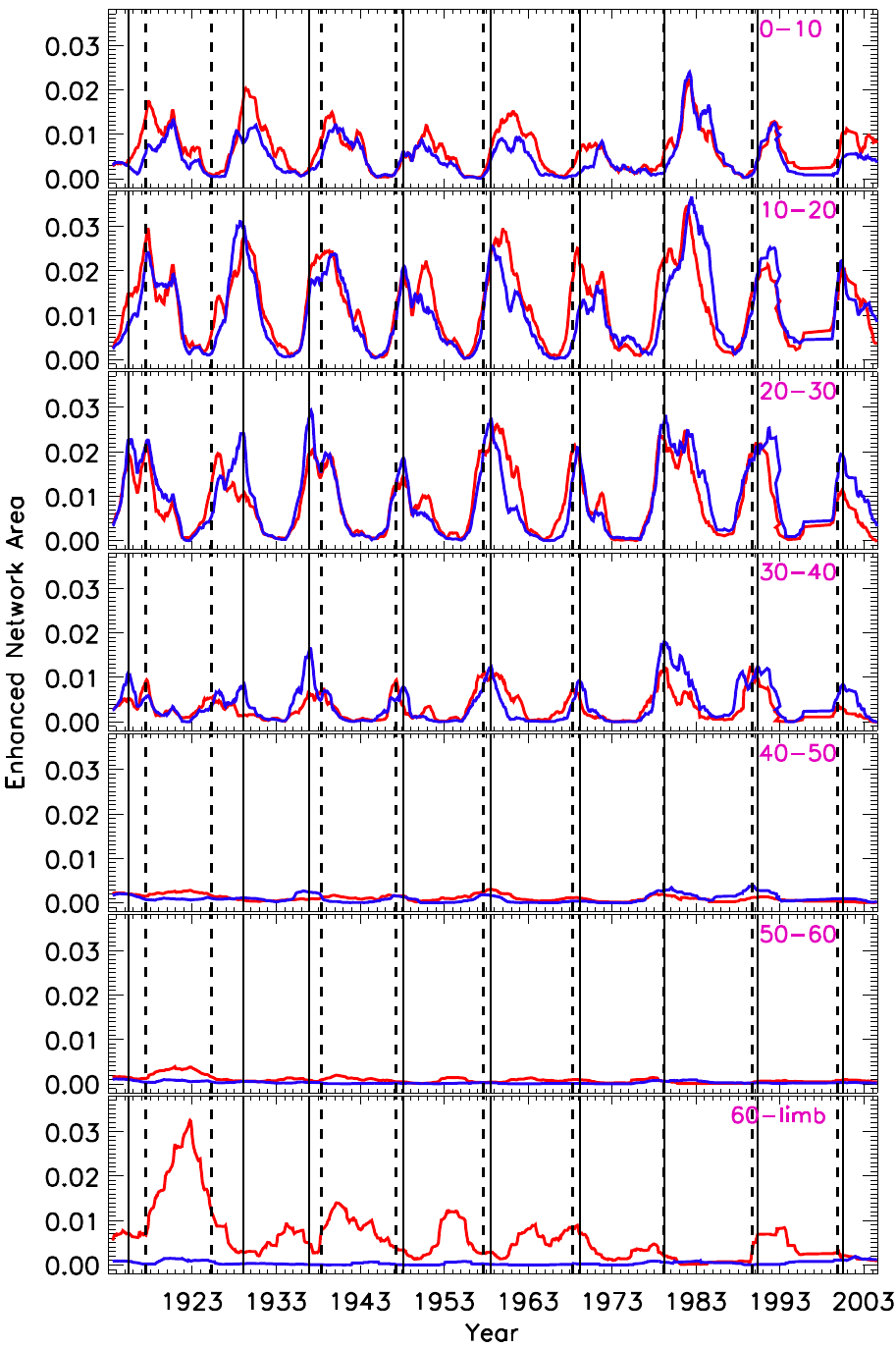}
\caption{Same as Figure \ref{p_north} but for EN area.
}
\label{en_north}
\end{figure*}

\subsubsection{Variation in Enhanced Network (EN) Area}
The Ca-K plages fragments into smaller regions when decay after magnetic field of the region weakens. The EN represents the decaying plages with average intensity less than that of plage regions \citep{Priyal2017}. To identify the EN, the threshold intensity is the same as that for plages but area limit is much smaller than that for plages. Six upper panels of Figure \ref{en_north} show the variation of monthly averaged fractional EN areas for latitude belts up to 60$\degree$ at an interval of 10$\degree$ and the bottom panel indicates the monthly averaged EN area for the latitude belt of 60$\degree$ -- limb for the both the hemispheres by red (north) and blue (south) curves. The fractional EN areas are less as compared to fractional plages areas for each latitude belt. The values of fractional areas around maximum phase of each cycle are more for 10 -- 20$\degree$ latitude belt as compared to other equatorial latitude belts. The average value of fractional EN area decreases from mid-latitude towards polar regions becomes minimum around 60$\degree$ latitude belts. The figure indicates that EN fractional areas are more in the northern hemisphere as compared to those in the southern hemisphere after 60$\degree$ latitude. It is not clear why it is so? It may be some instrumental effect but long term variations do exist. It may be noted that occurrence of EN areas is significant in the polar region whereas in the plage areas, it is very less. Further, the variation in EN areas as function of time in polar regions indicate cyclic variation with period of about 11-year and out of phase with the variations at 30 -- 40$\degree$ latitude belt. It will be verified using cross-correlation function. 

\begin{figure*}
\centering
\includegraphics[width=10cm]{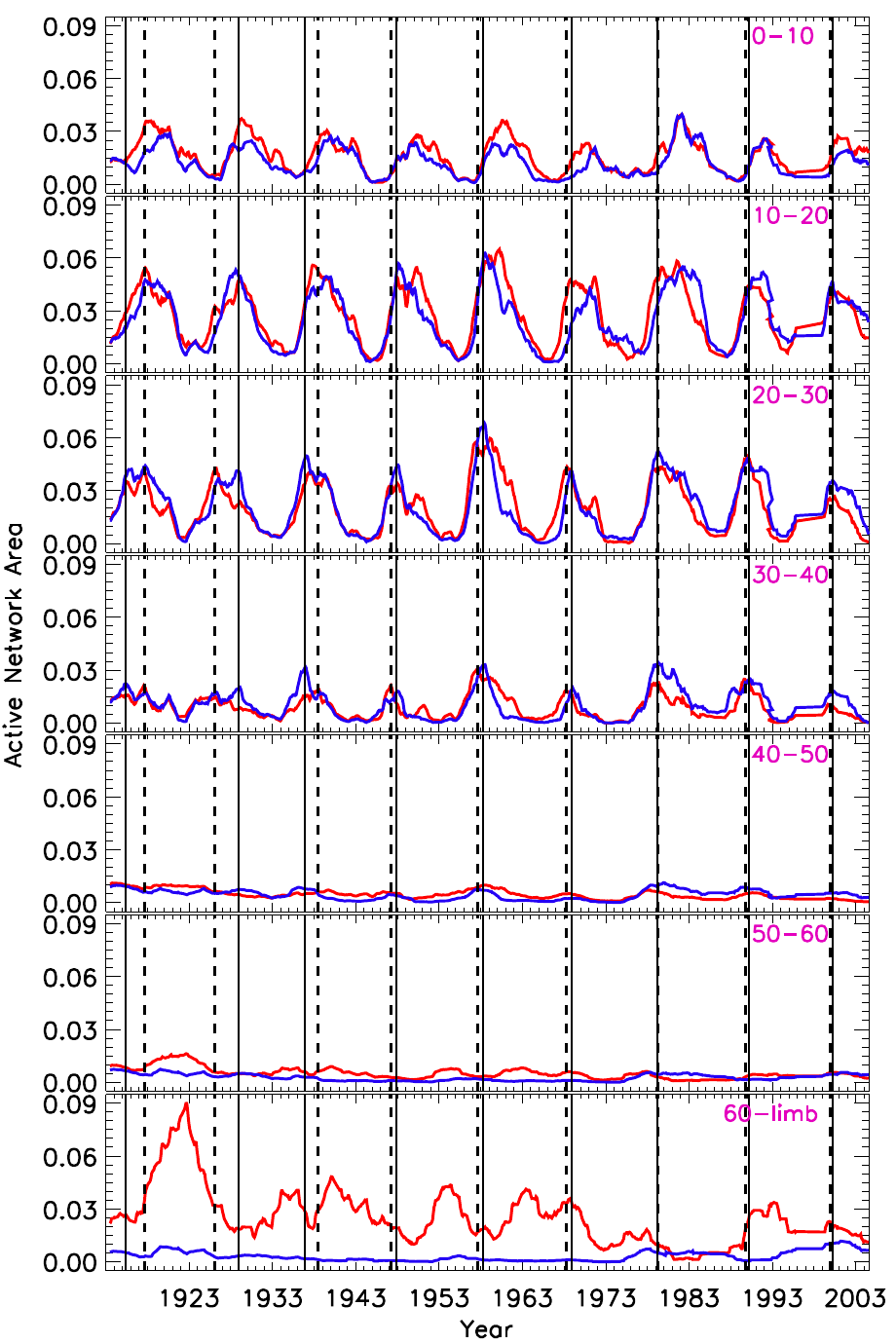}
\caption{Same as Figure \ref{p_north} but for AN area.
}
\label{an_north}
\end{figure*}

\subsubsection{ Variation in Active Network (AN) Area}
The AN areas are less bright than the EN areas and probably represent magnetic elements in the lanes between granulation. These can also be associated with decaying plages or some magnetic elements seen at the photospheric layer. The description of Figure \ref{an_north} is similar to that of Figure \ref{en_north} but for AN region. The variations of AN fractional area are similar to those EN fractional areas. The fractional areas of AN are more than that of EN fractional area, by about a factor of 2. The Solar Cycle variations in the AN fractional area are clearly visible in Figure \ref{an_north}, up to 40$\degree$ latitude belts throughout the period of Solar Cycles 15 -- 23 in both the hemispheres. But, the cyclic variations are seen up to 1974 only (Solar Cycle 15 -- 20) in the higher latitude belts. This may be because the data available for less number days year$^{-1}$ with frequent gaps after 1974. 
The amplitude variation in the fractional area of AN is similar for respective latitude belts in northern and southern hemisphere up to the mid-latitudes. This variation becomes very small in the latitudes from 40 -- 60$\degree$. At the poles ($>$ 60$\degree$), this variation in AN area is enhanced in northern hemisphere whereas, in southern hemisphere, this enhancement is not seen. There appears anti-correlation between the variations in the mid-latitudes and those in the polar latitude belts.
This will be confirmed by computing the cross-correlations functions between 30 -- 40$\degree$ latitude belt with other latitude belts.

\begin{figure}
\centering
\includegraphics[width=10cm]{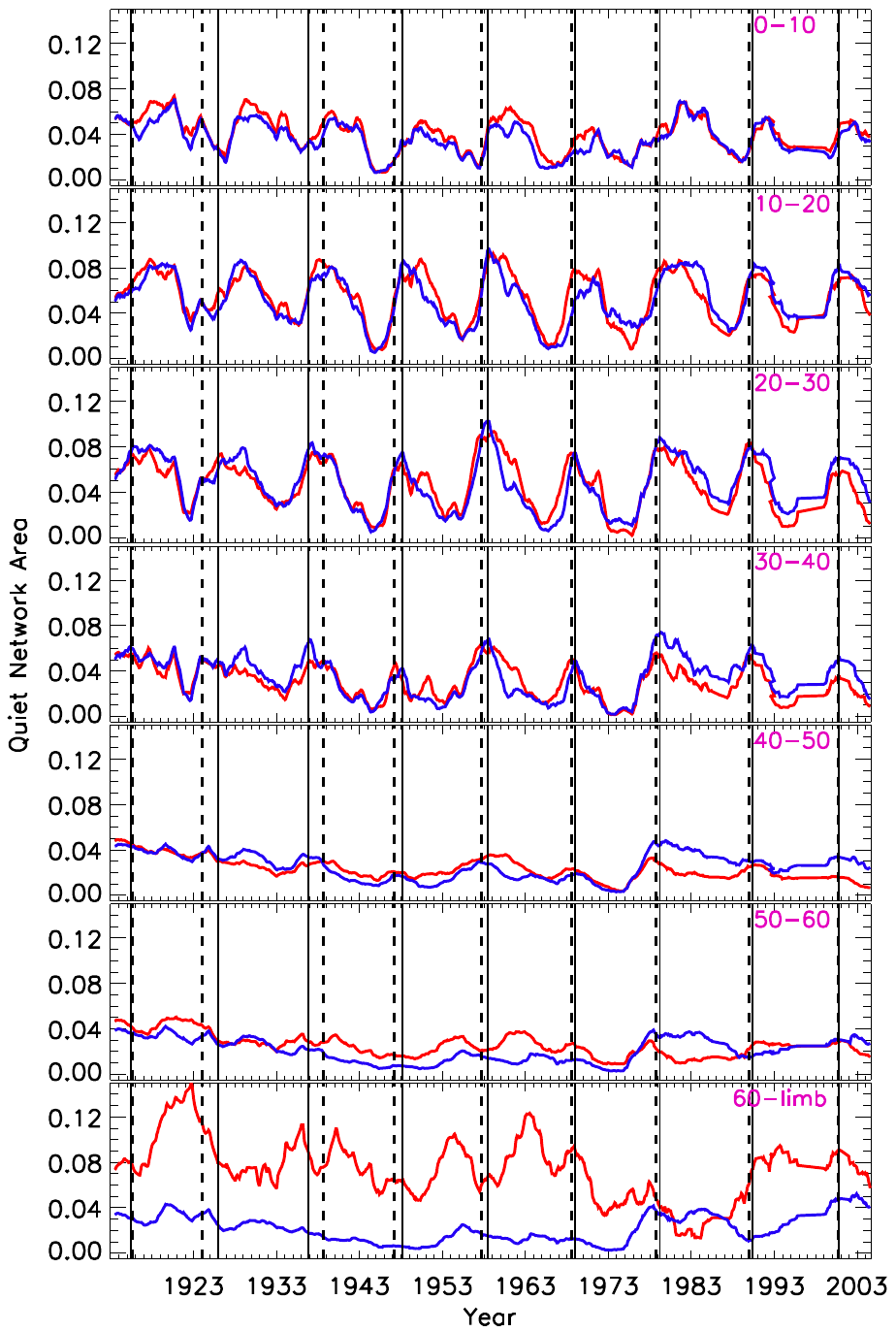}
\caption{Same as Figure \ref{p_north} but for QN area.}
\label{qn_north}
\end{figure}

\subsubsection{Variation in Quiet Network (QN) Area}
The QN regions are generally found at the boundaries of the large convective cells, known as super-granules. The horizontal flow from the center of the cell caries the flux tube and deposit at the boundary of the cell creating the increase in magnetic field at the boundary. This causes the increase in brightness of those regions and appearance of QN. In Figure \ref{qn_north} we show the monthly averaged fractional QN areas as a function of time 0 -- 60$\degree$ latitudes at an interval of 10$\degree$ for both the hemispheres in the upper six panels. The bottom panel of the figure shows the same for the 60$\degree$ -- limb. The figure indicate that QN fractional area varies with 11-year Solar Cycle up to 40$\degree$ latitude in both the hemispheres but at the higher latitude belts this cyclic variation is not seen clearly. Even though the QN is not directly related with solar activity but it appears that it plays some role in increasing magnetic field at the boundaries of super-granules. This causes increase in the fractional area of QN during the active phase of the Sun.   

\subsubsection{Comparision of Amplitude of Variation at Various Latitude Belts}
We have studied the variation of plage, EN, AN and QN areas as a function of latitude and time for Solar Cycles 15 -- 23. In Table \ref{table1}, we list the average values of amplitude of variations for plages, EN, AN and QN areas for various latitude belts. In case of fractional plage area the maximum amplitude of variation 
is 0.021 for 10 -- 20$\degree$ latitude belt and minimum amplitude 0.0002 for 50 -- 60$\degree$ latitude belt for the northern hemisphere. The values of amplitude of fractional plage area for the southern hemisphere also show the same trend as a function of latitude. The amplitude values in the table indicate that the fractional area of EN, AN and QN follows the similar trend as a function of latitude. Thus, it can be said that variations in the fractional area of plage, EN, AN and QN with time are maximum around 20$\degree$ and minimum around 50$\degree$ latitude belts which agree the findings of \inlinecite{Sindhuja2014}. They reported that there is minimum variation in the Ca-K$_1$ width around 50$\degree$ latitude belt. 

\begin{center}
\tabcolsep=0.5cm
\begin{table}[ht!]
\renewcommand{\arraystretch}{1.7}
\begin{tabular*} {\textwidth} {ccccc}
\cline{1-5}
Latitude Belt &  \multicolumn{4}{c}{Average Amplitude Variation of Chromospheric Features} \\
\cline{2-5}
 & Plage & EN & AN & QN \\
\cline{1-5}
0 -- 10 N	 & 0.0101 & 0.0065 & 0.0162 & 0.0403 \\
10 -- 20 N	 & 0.0210 & 0.0121 & 0.0279 & 0.0549 \\
20 -- 30 N	 & 0.0114 & 0.0087 & 0.0204 & 0.0469 \\
30 -- 40 N	 & 0.0024 & 0.0032 & 0.0092 & 0.0314 \\
40 -- 50 N 	 & 0.0004 & 0.0011 & 0.0050 & 0.0237 \\
50 -- 60 N 	 & 0.0002 & 0.0010 & 0.0054 & 0.0257 \\
60 -- limb N	 & 0.0014 & 0.0064 & 0.0262 & 0.0741 \\
0 -- 10 S	 & 0.0063 & 0.0047 & 0.0126 & 0.0361 \\
10 -- 20 S	 & 0.0189 & 0.0108 & 0.0255 & 0.0529 \\
20 -- 30 S 	 & 0.0119 & 0.0094 & 0.0211 & 0.0482 \\
30 -- 40 S	 & 0.0030 & 0.0039 & 0.0103 & 0.0345 \\
40 -- 50 S	 & 0.0004 & 0.0009 & 0.0045 & 0.0254 \\
50 -- 60 S	 & 0.0000 & 0.0003 & 0.0027 & 0.0206 \\
60 -- limb S	 & 0.0001 & 0.0004 & 0.0030 & 0.0210 \\
\cline{1-5}
\end{tabular*}
\caption{Table showing the average amplitude variation of fractional areas of chromospheric features (plage, EN, AN, and QN) in different latitude belts. ``N'' and ``S'' are used for northern and southern hemispheres, respectively.}
\label{table1}
\end{table}
\end{center}
\vspace{-1cm}

These findings do not support the \inlinecite{Babcock1961} model which says that toroidal part of magnetic field moves towards the equator and poloidal component moves towards pole from mid-latitude belts. One can ask why there is minimum activity around 50$\degree$ latitude belts as compared to mid latitude belts and polar region if weak magnetic field moves towards polar regions from mid latitude belts? Further, \inlinecite{Raouafi2007} reported that the magnetic field elements remain approximately same between 55$\degree$ and 75$\degree$ and afterwards they decrease by more than 50$\%$ toward the pole. But, \inlinecite{Jin2012} restricted their analysis of the Solar Cycle variation of the magnetic field only up to 60$\degree$ because they found fewer magnetic elements and more noise while approaching the polar regions. The cross-correlation curves of EN and AN do indicate the significance existence of small scale magnetic elements at the polar region.

\subsection{Phase Difference Analysis for the Latitude Belts}
The activity diagram, popularly known as butterfly diagram indicates that the sunspots and related activity such as plages appear at middle latitudes around 40$\degree$ at the beginning of Solar Cycle. The sunspots and plages remain visible from few days to few months and then decay by fragmentation as the related magnetic field weakens. These appear at lower and lower latitudes as the Solar Cycle progresses. The speed of movement can be determined from the phase differences between the two latitude belts considered. Therefore, to determine movement of activity we have computed the cross-correlation function (CC) for various latitude belts with 30 -- 40$\degree$ latitude belt as the Solar Cycle activity begins in this belt. Generally, it is advisable to investigate the meridional flow speed for each cycle considering the data for about 20 years. But to begin with, we consider the data for the whole period of 1913 -- 2004 to determine the phase difference of the occurrence maximum activity at different latitude belts with respect to the mean latitude of 35$\degree$. To compute the cross-correlation function we use monthly averages of plage, EN, AN and QN areas.

\begin{figure*}
\centering
\includegraphics[width=\textwidth]{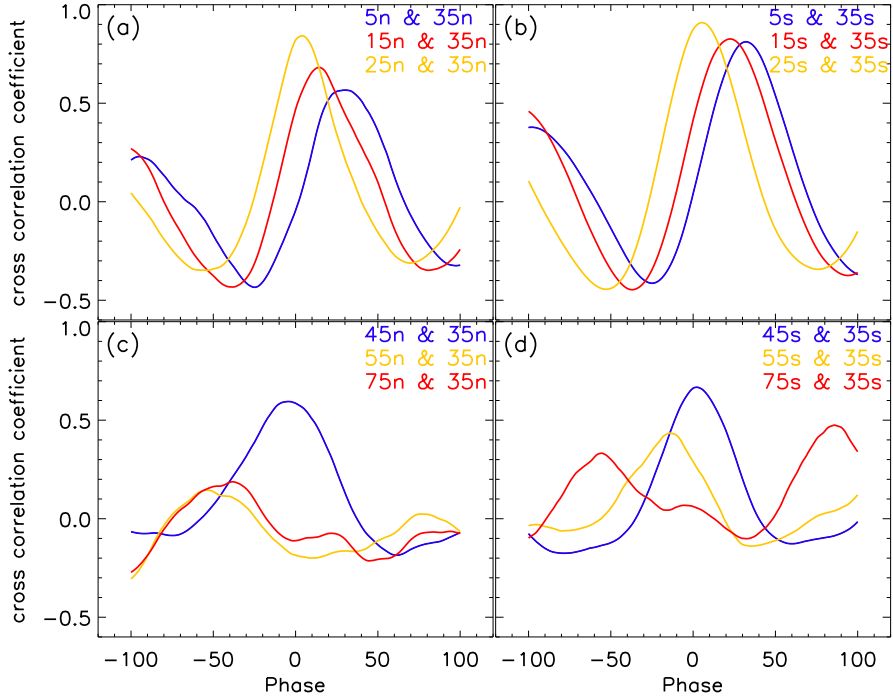}
\vspace{-6.8cm}
\caption{Plots of cross-correlation coefficient of various latitude belts with 35$\degree$ latitude belt as a function of phase lag in months for plage area considering the data for the period 1913 -- 2004, Solar Cycles 15 -- 23. Panel (a) and (b) displays the plots for 5$\degree$ (blue), 15$\degree$ (red) and 25$\degree$ (yellow) with 35$\degree$ latitude belt for northern and southern hemisphere,respectively. Panel (c) and (d) shows the plots for 45$\degree$ (blue), 55$\degree$ (yellow), and 75$\degree$ (red) for northern and southern hemisphere, respectively.}
\label{corr_plage}
\end{figure*}

\subsubsection{Cross-correlation Functions for Plages}
As mentioned above, we have determined the plage, EN, AN and QN areas at an interval of 10$\degree$ up to 60$\degree$ latitude and from 60$\degree$ to visible latitude in both the hemispheres on daily basis. We refer each latitude belt by its mean latitude, for example 5n for 0 -- 10$\degree$ north latitude belt and 5s for 0 -- 10$\degree$ south latitude belt. The computed values of cross-correlation function (CC) for various latitude belts with respect to 35$\degree$ latitude belt as a function of phase lag in months for plage areas are plotted in Figure \ref{corr_plage}. In the top left panel of the figure we plot CC values of 5$\degree$, 15$\degree$ and 25$\degree$ latitude belt with 35$\degree$ belt up to a phase lag of 100 months for the northern hemisphere shown in blue, red and yellow curves, respectively. The right top panel shows the same for the southern hemisphere. The CC values vary smoothly with the phase lag and maximum values of CC are very good indicating that accurate value of phase lag can be determined. Bottom row two panels show the CC curves for the higher latitudes of 45$\degree$, 55$\degree$ and 75$\degree$ with 35$\degree$ belt (hereafter simply referred as 45, 55 75, etc.)  for both the hemispheres by blue, yellow and red curves, respectively. The maximum values of CC (hereafter called CC-max) are very low as negligible number of plage occur at high latitude belts. The negative value of the phase difference between 35n with 45n indicate that the activity began early at the higher than 35$\degree$ latitude belt. It may be due to coincidence as the plage activity in 45$\degree$ belt is very low. On the other-hand the phase difference is positive for the 35s with 45s belt. This indicate that either some of the fragmented plages moved towards polar region or coincidence because of less plage activity in 45$\degree$ belt. This needs to be investigated further considering the data for one to two consecutive Solar Cycles. The values of CC-max and phase difference for plages for different latitude belts are listed in Table \ref{table2}. The CC curves for the 55 and 75 belts show negligible correlation or anti-correlation for zero phase lag. The CC curves in northern hemisphere for 55 and 75 do not show phase lag clearly whereas 55 belt indicate in the southern hemisphere indicate a phase lag of about 14 months. And 75s belt shows negligible correlation or anti-correlation at zero phase lag but two peaks separated by about 11-years, one each side of the zero phase lag. This type of behaviour may be due to the reason that plages are rarely visible at higher latitudes. 

\begin{figure*}
\centering
\includegraphics[width=\textwidth]{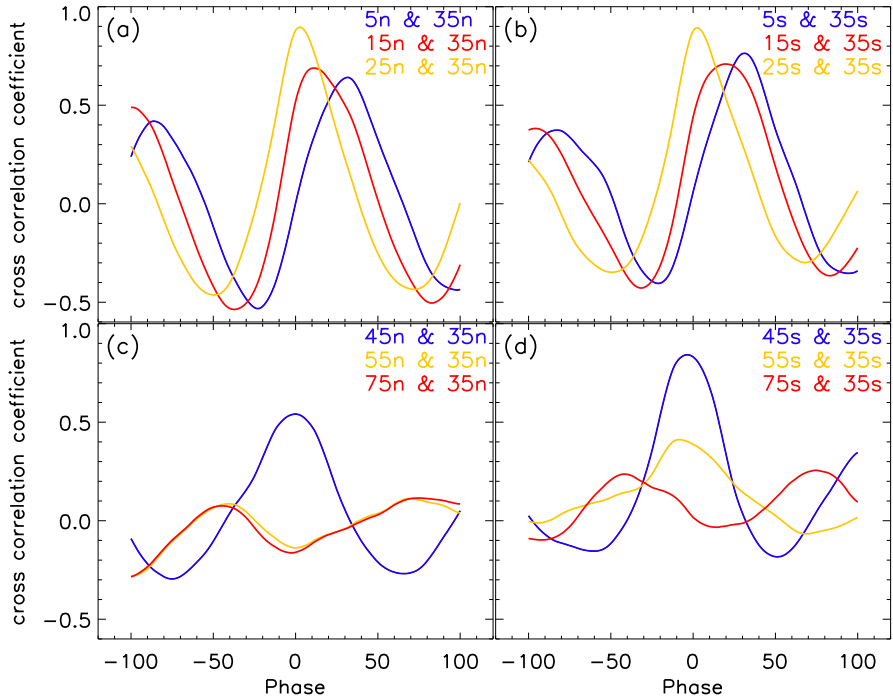}
\vspace{-6.8cm}
\caption{Same as Figure \ref{corr_plage}, but for EN area.}
\label{corr_enhance}
\end{figure*}

\subsubsection{Cross-correlation Functions for Networks}
\begin{figure*}
\centering
\includegraphics[width=\textwidth]{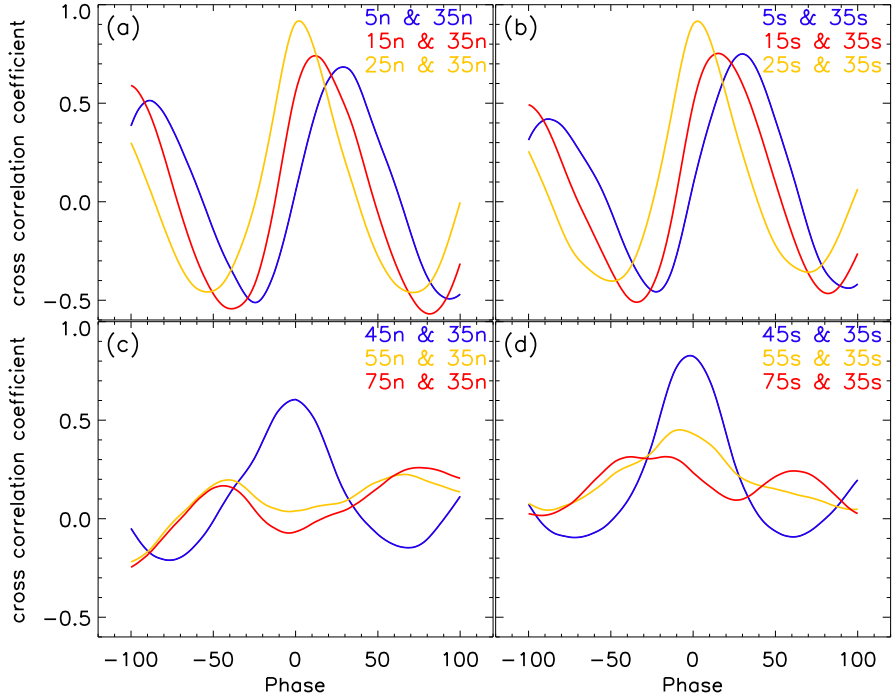}
\vspace{-6.8cm}
\caption{ Same as Figure \ref{corr_plage}, but for AN area.}
\label{corr_active}
\end{figure*}

\begin{figure*}
\centering
\includegraphics[width=\textwidth]{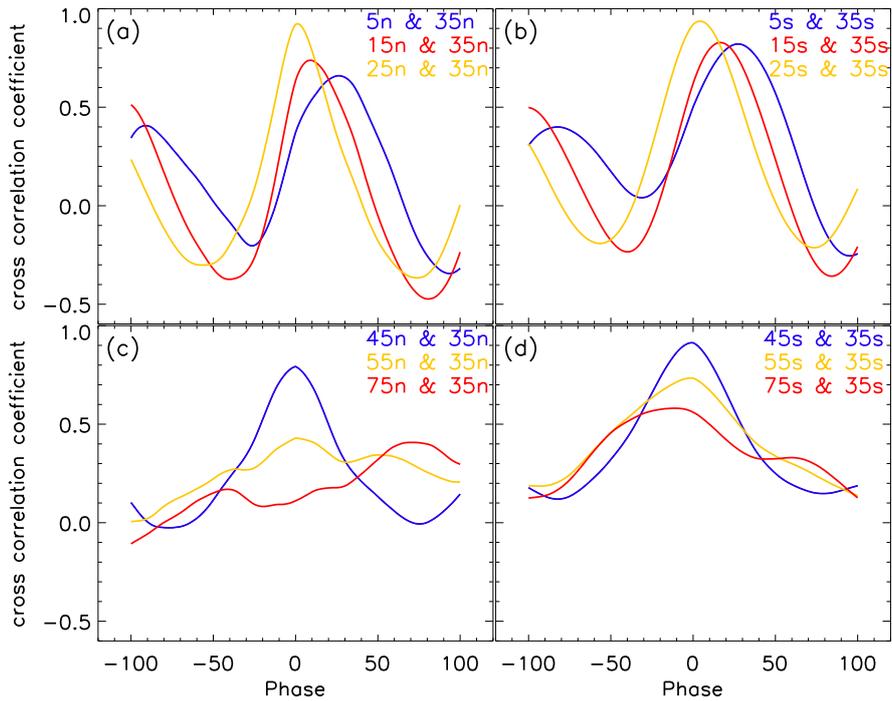}
\vspace{-6.8cm}
\caption{Same as Figure \ref{corr_plage}, but for QN area. }
\label{corr_network}
\end{figure*}

In the top two panels of Figure \ref{corr_enhance} we plot the CC values for the 5, 15 and 25 latitude belts as a function of phase lag in months for the northern hemisphere (left panel) and southern hemisphere (right panel) for the EN. The CC curves and phase lags are similar to those for the plage areas. The curves for the northern hemisphere indicate a phase lag of about 3 months for occurrence of maximum activity between 25$\degree$ and 35$\degree$ latitude belt. The phase becomes relatively larger about 11 months for 15$\degree$ and 35$\degree$ belts. Then the phase lag increases at a faster rate for 5$\degree$ and 35$\degree$ latitude belts. This behaviour indicates that activity does not shift from middle latitude belts towards equator at a uniform speed. It implies that at the beginning activity shift at a faster rate and later it moves at a lower rate towards equator. Bottom two panels of Figure \ref{corr_enhance} show the CC curves for the higher latitude belts, namely 45, 55 and 75 belts. The corresponding curves are similar to those for plage areas seen in Figure \ref{corr_plage}. Similarly, the Figures \ref{corr_active} and \ref{corr_network} show the CC curves for the AN and QN, respectively.

The Figures \ref{corr_plage} -- \ref{corr_network} indicate that CC curves are similar for the equatorial latitude belts in case of plage, EN, AN and QN areas with similar phase difference between respective latitude belts. The values of phase difference for each pair of latitude belt are given in Table \ref{table2}. But, the CC curves differ for higher latitude belts especially for 55 and 75 belts. The CC curves for 55 and 75 latitude belts are similar for EN and AN. The CC curve for 55n shows double peak around the zero phase lag whereas it is not seen in CC curve for 55s belt. But both the 75n and 75s belts show double peak separated by about 11 years indicating anti-correlation in the occurrence maximum activity between equatorial and polar regions. 
The CC curves for 55n, 55s and 75s belts for QN show a single peak with very small phase lag with corresponding 35$\degree$ latitude belt. Whereas, the 75n latitude belt shows double peak but of not significance separated by about 11 years. The area of QN at the equatorial belts do vary with about 11-year periodicity, probably due to the effect of decaying active regions
In general, we can say that activity in the polar regions is anti-correlated with the activity at the equatorial regions even though the values of CC-max are very low.

%%%%%%%%%%%%%%%%%%%%%%%%%%%%%%%%%%%%%%%%%%%%%%%%%%%%%%%%%%%%%%%%%%%%%%%%%%%%%%%%%%%%%%
\section{Summary}
\label{sect_Summary}
Most of the people believe that dynamo process operating at the base of convection zone generates magnetic field. This leads to formation of sunspots, plages, filaments and other active features on the solar surface. The systematic variations in these features with time have been used to study two components of magnetic field, namely toroidal and poloidal \citep{Choudhuri1995}. The occurrence of large active features such as sunspots and plages shift towards equator from middle latitudes and small active features like network elements move towards polar region. This shift is caused by flow of material in the meridional plane from the solar equator toward the Sun's poles and from the poles toward the equator deep inside the
Sun \citep{Choudhuri1995, charbonneau2007}. The meridional flow plays an important role in the Sun's magnetic dynamo to carry the dynamo wave toward the equator. \inlinecite{Howard1981} reported that the polar fields are formed due to the migration of weak
magnetic field toward the pole. The observed weak fields at polar regions are due to the poloidal field generated from the solar dynamo process and the pole-ward migration of small scale
magnetic fields of decaying solar active regions. The aforementioned processes are likely to cause the reversal of the polarity of the poloidal field at the polar region \citep{Leighton1964, makarov1999}. Dynamo models suggest that meridional flow towards the equator plays an important role to transport the toroidal component of the magnetic field \citep{Hathaway2003}. The magnetic field data of the Sun has been used to study the variations of the magnetic field with the Solar Cycle \citep{Jin2012}. But the detailed magnetic field images have been obtained from the last part of the 20$^{th}$ century. 

The chromospheric such as Ca-K images of the Sun can be used as proxy to study the magnetic fields on the Sun which are available from the beginning of the 20$^{th}$ century. The Ca-K plages represent strong toroidal component of magnetic field whereas networks such as EN, AN and QN indicate weak poloidal fields.

\inlinecite{Hathaway2011} found that the meridional flow is faster at minimum as compared to that at maximum phase. Further, the meridional flow speed during the initial phase of Solar Cycle 23/24 minimum was substantially faster than that at the Cycle 22/23 minimum. The average latitudinal profile is a sinusoidal that extends to the poles and peaks at about 35$\degree$ latitude. With the progress in the Solar Cycle, a pattern of inflows toward the sunspot zones develops and moves equator-ward as the sunspot zone shifts towards equator. They found a peak pole-ward meridional flow velocity of 11.2 \ms at a latitude of 35$\degree$ and an average meridional flow profile substantially different in the northern and southern hemispheres. The reported faster flow velocity in the southern hemisphere with peak at a higher latitude than in the northern hemisphere. Their measurements indicated that flows almost vanish at the extreme northern limit (75$\degree$) while the pole-ward flow with a speed of about 5 \ms at the southern limit persists. \inlinecite{Hathaway2014} find that the systematic weakening of the meridional flow on the pole-ward sides of the active (sunspot) latitudes. They interpreted this as an inflow toward the sunspot zones superimposed on a general pole-ward meridional flow profile. They also found that the meridional flow varied from cycle to cycle. Models of the magnetic flux transport by a variable meridional flow suggest significant modulation on the size and timing of the following sunspot cycle through its impact on the Sun's polar magnetic fields. 

 \inlinecite{Imada2018} found that meridional circulation velocity peaked at $\approx$ 12 \ms at a latitude of 45$\degree$. Their measurements showed that the magnetic elements with stronger and weaker magnetic fields largely represent the characteristics of the active region remnants and solar magnetic networks, respectively. They found that magnetic elements with a strong (weak) magnetic field show a faster (slower) rotation speed. On the other hand, magnetic elements with a strong (weak) magnetic field show slower (faster) meridional circulation velocity. \inlinecite{Imada2020} found that the average meridional flow profile peaked at $\approx$15 \ms at 45$\degree$. During the declining phase of the cycle, the meridional flow at the middle latitude (30$\degree$) accelerated from 10 to 17 \ms in both hemispheres. All these measurements indicate that meridional flows peaks around 40$\degree$ latitude.

\begin{center}
\tabcolsep=0.07cm
\begin{table}[ht!]
	\renewcommand{\arraystretch}{1.8}

\begin{tabular*} {\textwidth} {ccccccccc}
\cline{1-9}
Correlation &  \multicolumn{4}{c}{Maximum Correlation Coefficient} & \multicolumn{4}{c}{Phase Difference in Months} \\
\cline{2-9}
Between  & Plage & EN & AN & QN &  Plage & EN & AN & QN  \\
Latitude  &  &  &  & &  & & &   \\
                \cline{1-9}
5n and 35n     & 0.57        & 0.64      & 0.68        & 0.66         & 30            & 32     & 29         & 26.5            \\
15n and 35n    & 0.68        & 0.69     & 0.74        & 0.74         & 14.5            & 11        & 12          & 8.5             \\
25n and 35n    & 0.84        & 0.89     & 0.92        & 0.92         & 3.5             & 2.5       & 2.5             & 1.5             \\
35n and 35n    & 1             & 1     & 1          & 1             & 0             & 0             & 0             & 0             \\
45n and 35n    & 0.59        & 0.54     & 0.60       & 0.79         & -4.5            & -0.5            & -0.5            & -0.5            \\
55n and 35n    & 0.15, 0.02  & 0.09, 0.11 & 0.19, 0.22  & 0.43  & -54, 76       & -41,69   & -41, 66       & -0.5         \\
75n and 35n    & 0.19, -0.06 & 0.07, 0.11 & 0.17, 0.26  & 0.17, 0.41  & -38, 94       & -45, 75     & -44, 75.5         & --       \\
5s and 35s     & 0.81        & 0.76     & 0.75         & 0.82         & 32           & 31.5     & 29            & 27            \\
15s and 35s    & 0.83        & 0.71     & 0.75         & 0.83         & 22.5           & 19.5   & 15            & 17            \\
25s and 35s    & 0.91        & 0.89     & 0.92        & 0.94         & 5.5         &2.5    & 3         & 4             \\
35s and 35s    & 1             & 1     & 1             & 1             & 0             & 0             & 0             & 0             \\
45s and 35s    & 0.67        & 0.84     & 0.83        & 0.91         & 2           & -3.5   & -2          & -1            \\
55s and 35s    & 0.43        & 0.41     & 0.45       & 0.73         & -14           & -7    & -8.5        & -1.5            \\
75s and 35s    & 0.33, 0.47  & 0.24, 0.26  & 0.31, 0.24 & 0.58, 0.33  & -56, 86      & -41.5, 74    & -17, 61        & --        \\

\cline{1-9}
\end{tabular*}
\caption{Table of Maximum correlation coefficient with phase difference for different latitudes with 35$^{\circ}$ latitude. The maximum correlation is for different features (plage, EN, AN, QN) from Solar Cycle 15 -- 23.}
\label{table2}
\end{table}
\end{center}
\vspace{-1cm}

%%%%%%%%%%%%%%%%%%%%%%%%%%%%%%%%%%%%%%%%%%%%%%%%%%%%%%%%%%%%%%%%%%%%%%%%%%%%%%%%
The phase difference between different latitude belts may vary from cycle to cycle in both the hemispheres. Here we are considering average values of phase difference for the period of 1915 -- 2004. The values of phase differences for the plages, EN, AN and QN are similar for the northern hemisphere for different latitude belts.
Table \ref{table2} indicates that it takes about 30 months for the toroidal field to travel from mid-latitudes to equator in both the hemispheres but total time may be larger than 30 months as this period is for 35$\degree$ to 5$\degree$ latitudes only. There may be overall symmetry in the northern and southern hemispheres but not at shorter time scales as indicated by the values of phase lag between different latitude belts in the northern and southern hemispheres. The values of phase lag for different latitude belts indicate that the variations in plage areas, EN, AN and QN are in-phase with each other with some differences. It is difficult to comment on these differences without the detailed analysis on shorter time scales. The scatter and gaps in the data results significant variations in values of phase lag on shorter time scales. The average phase difference is $\approx$ 3 months in the occurrence of maximum activity $\approx$ 35$\degree$ and 25$\degree$ latitude belts in the northern hemisphere. This value of phase lag indicates a speed of the activity shift is $\approx$ 19 \ms at the beginning Solar Cycle in these two belts, from higher to lower latitude in the northern hemisphere. Then speed slowed down to 5.4 \ms between 25 to 15$\degree$ belts and further slowed to 2.7 \ms near the end phase of cycle. But in the southern hemisphere these values are $\approx$ 11.8, 2.8 and 5.4 \ms for the corresponding latitude belts. These values imply that shift in activity from mid to lower latitude in northern and southern hemispheres is not exactly symmetric. At the same time these values indicate that the activity shifts at a faster rate at the beginning of the Solar Cycle in higher to mid latitude belts and then at a slower rate in lower latitude belts as the cycle progresses. These findings of maximum speed occurring around 35$\degree$ and an average amplitude of about 15 \ms agree with those of \inlinecite{Hathaway2011}, \inlinecite{Imada2020} and others. The differences in the shift of activity may be due to latitude dependence of flux emergence, cycle variations, difference in meridional flow velocities in north-south hemispheres.

These findings need to be confirmed by using data of individual Solar Cycle and of better quality. Due to scatter in the data we had to take averages over longer period of time. The shift of the poloidal field towards polar region is not clear as the activity around 50$\degree$ belt is very low and scatter in the data. But it is clear that activity near polar region around 75$\degree$ is anti-correlated with the activity at mid-latitudes. We have shown that it is possible to study the meridional flows using the Ca-K images and analysing the data as function of latitude and time. The results of this study will be very useful to understand dynamics of the Sun.

\vspace{0.8cm}

\noindent {\bf Disclosure of Potential Conflicts of Interest}  The authors declare that they have no conflicts of interest.
\begin{acks}
	We would like to thank the referee for the valuable comments.
We thank the numerous observers who has made the observations over a century and kept the data in good environment conditions. We acknowledge the enormous work done by the digitization team lead by Jagdev Singh.
PD thanks the CSIR, New Delhi for their support. RJ acknowledges the Department of Science and Technology (DST), Government of India for the INSPIRE fellowship.
\end{acks}

%----------------------------------
\mbox{}~\\
\bibliographystyle{spr-mp-sola}
\bibliography{reference}
\IfFileExists{\jobname.bbl}{} {\typeout{}
\typeout{***************************************************************}
\typeout{***************************************************************}
\typeout{** Please run "bibtex \jobname" to obtain the bibliography}
\typeout{** and re-run "latex \jobname" twice to fix references}
\typeout{***************************************************************}
\typeout{***************************************************************}
\typeout{}}
%\section{Acknowledgments}
\end{article} 
\end{document}